\documentclass[12pt,a4paper,superscriptaddress]{revtex4}
\usepackage{graphicx}
\usepackage{graphics}
\usepackage{amsmath}
\usepackage{dcolumn}
\usepackage{amssymb}
\usepackage{bm}
\usepackage[latin1]{inputenc}
\newcommand{\be}{\begin{equation}}
\newcommand{\ee}{\end{equation}}
\newcommand{\bea}{\begin{eqnarray}}
\newcommand{\eea}{\end{eqnarray}}

\newcommand{\pa}{\partial}

\begin{document}

\title{On the effective potential, Horava-Lifshitz-like theories and
  finite temperature}

\author{C. F. Farias}

\affiliation{Departamento de F\'{\i}sica, Universidade Federal da 
Maranh\~{a}o,\\
65085-580, S\~ao Luiz, MA, Brazil}
\email{cffarias@gmail.com}

\author{M. Gomes}

\affiliation{Departamento de F\'{\i}sica Matem\'{a}tica,
Instituto de F\'{\i}sica,
Universidade de S\~{a}o Paulo\\
05315-970, S\~ao Paulo, SP, Brazil}
\email{mgomes,ajsilva@fma.if.usp.br}

\author{J. R. Nascimento}

\affiliation{Departamento de F\'{\i}sica, Universidade Federal da
Para\'{\i}ba\\
Caixa Postal 5008, 58051-970, Jo\~ao Pessoa, PB, Brazil}
\email{jroberto,petrov@fisica.ufpb.br}

\author{A. Yu. Petrov}

\affiliation{Departamento de F\'{\i}sica, Universidade Federal da 
Para\'{\i}ba\\
Caixa Postal 5008, 58051-970, Jo\~ao Pessoa, PB, Brazil}
\email{jroberto,petrov@fisica.ufpb.br}

\author{A. J. da Silva}

\affiliation{Departamento de F\'{\i}sica Matem\'{a}tica, 
Instituto de F\'{\i}sica,
Universidade de S\~{a}o Paulo\\
05315-970, S\~ao Paulo, SP, Brazil}
\email{mgomes,ajsilva@fma.if.usp.br}

\begin{abstract}
 We calculate the one-loop effective
potential at finite temperature for the
Horava-Lifshitz-like QED and Yukawa-like theories for arbitrary values
of the critical  exponent and the space-time dimension. Additional
remarks on  the
zero temperature situation are also presented.
\end{abstract}
\maketitle
 
\section{Introduction}
The Horava-Lifshitz (HL) methodology, based
on an asymmetry between time and space coordinates
\cite{Hor}, has gained much attention within the context of the
search for a perturbatively consistent gravity theory. The main
advantage of that approach comes from the fact that, from one side, it
improves the renormalizability of field theory models, and, from
another side, it avoids the appearance of ghosts whose presence is
characteristic of theories with higher time derivatives
\cite{Hawk}. Therefore, this concept (or, more generally, the concept
of time-space asymmetry) began to be applied not only within
studies of gravity but also for  other
(f.e. scalar and vector) field theory  models. 

One line of studies of  theories with  time-space asymmetry is
devoted to the investigation of their renormalization. Within this
context, the HL versions of the gauge
field theories \cite{ed}, scalar field theories \cite{Anselmi},
four-fermion theory \cite{ff} and $CP^{N-1}$ model \cite{cpn} were
considered. Another important result in this context is the
generalization of the Ward identities for the HL-like theories  \cite{Gomes}.

Another line of investigations on  HL-like theories concerns  the study
of the effective potential. In the works
\cite{Eune,Liou,Fara,our} the one-loop effective potential for 
scalar field theories with different forms of self-couplings and
arbitrary values of the critical exponent $z$, for  scalar QED and
for the Yukawa models with $z=2$ and $z=3$ have been obtained. However,
a remaining problem was the calculation of the (one-loop) effective
potential for the same models with an arbitrary value
of the critical exponent. The analysis of these models at zero
temperature has been carried out in \cite{our1}, and its extension
for the non-zero temperature case is considered in this  paper.

We begin our study  of  scalar quantum electrodynamics at finite
temperature for generic $z$ 
and $d$ space dimensions by considering in the  section 2 that the
effective space-time dimension $d+z$ 
is odd. As we shall demonstrate, at these values of $d$ and $z$ no
self-interaction of the scalar field 
is necessary to achieve the consistency of the model. However, as
discussed  in  the section 3,
for $d+z$ even, a divergence occurs
demanding the inclusion
of a self-interaction of the scalar field. 
In the section 4 we analyze an HL version of the Yukawa model and show
that for odd $z$ the model is in general nonrenormalizable with  only
one  exception happpening if the  needed counterterm is proportional
to  $\phi^{4}$.
  A summary and further comments of our results are
presented in section 5. 

\section{Scalar quantum electrodynamics  with $d+z$ odd} 
The Lagrangian of the scalar QED  with an arbitrary $z$ looks like 
\bea
\label{lasqed}
L&=&\frac{1}{2}F_{0i}F_{0i}-\frac{1}{4}F_{ij}(-\Delta)^{z-1} F_{ij}+D_0\phi
(D_0\phi)^*-
D_{i_1}D_{i_2}\ldots D_{i_z}\phi(D_{i_1}D_{i_2}\ldots D_{i_z}\phi)^*.
\eea
where $D_0=\pa_0-ieA_0$, $D_i=\pa_i-ieA_i$ is a gauge covariant
derivative and  we assume $\phi$ to be massless, for simplicity. By the
same reason,   we choose the critical exponents
for the scalar and vector 
fields to be the same.

Adding the gauge fixing term \cite{our1}
\bea
L_{gf}=-\frac{1}{2}\left[(-\Delta)^{-(z-1)/2}\partial_0A_0-
(-\Delta)^{(z-1)/2}\partial_i A_i\right]^2,
\eea
the propagators acquire the simple forms
\bea
<A_0A_0>&=&-\frac{ik^{2z-2}}{k^2_0+k^{2z}},\nonumber\\
<A_iA_j>&=&\frac{i\delta_{ij}}{k^2_0+k^{2z}}.
\eea
For the one-loop calculation of the effective potential, the  only relevant vertices are
\bea
&&e^{2} A_{0}^{2}\Phi\Phi^{\ast}, \qquad -ie(\Phi^{\ast}\phi -\Phi \phi^{\ast})\partial_{0}A_{0},\nonumber\\
&& -ie(\Phi\phi^{\ast} -\Phi^{\ast} \phi)\partial_{j}(-\triangle)^{z-1}A_{j},\qquad -e^{2}A_{j}(-\triangle)^{z-1}A_{j}\Phi\Phi^{\ast}
\eea
where $\Phi$ is a constant background scalar field generated by the shift $\phi\rightarrow\phi+\Phi$.

 From now on, except where explicitly indicated, the propagators (as well as
all momenta) will be taken in
the Euclidean space, and everywhere $k^2\equiv \vec{k}^2=k_ik_i$, with $i$ running from 1 to $d$.
As we will show shortly, similarly to what happens in the relativistic QED, the
perturbative consistency of the model  may require the addition 
of a self-interaction term for the scalar field. For the time being,
we discard such possibility so that, as it has been shown 
in \cite{our1} and will be discussed in more details in the next
section, the effective potential turns out to be determined by
a single integral
\bea
U^{(1)}&=&U_a+U_b+U_c=
\frac{d}{2}\int\frac{d^dkdk_0}{(2\pi)^{d+1}}\ln[k^2_0+
k^{2z}+
M^2k^{2z-2}],
\eea
where $M^2=2e^2\Phi\Phi^*$ and $U_{a}$, $U_{b}$ and $U_{c}$  are the
contributions coming from loops  containing only
$<A_{0}A_{0}>$ or $<A_{i}A_{j}>$  propagators and graphs with both the gauge
and scalar  field propagators,
respectively.

In the finite temperature case, following the Matsubara
prescription \cite{Matsubara}, 
observing that  all propagators
are bosonic ones, we must change  $k_0\to 2\pi n T$,
where $T$ is the temperature, and $n$ is an integer number.
The integral over $k_0$ is then replaced by the sum:
\bea
U^{(1)}&=&\frac{d}{2}T\sum\limits_{n=-\infty}^{\infty}\int\frac{d^dk}{(2\pi)^d}
\ln[
4\pi^2n^2T^2
+k^{2z}+M^2k^{2z-2}].
\eea
Using the known summation formula  \cite{DJ}:
\bea
\label{sumbos}
\sum\limits_{n=-\infty}^{\infty}
\ln(4\pi^2n^2T^2+E^2)=\frac{E}{T} +2\ln(1-e^{-E/T})+{\rm const},
\eea
where the additive constant does not depend on $E$ and will be omitted
from now on,
we have
\bea
U^{(1)}&=&\frac{d}{2}\int\frac{d^dk}{(2\pi)^d}\left\{ (k^{2z}+M^2
k^{2z-2})^{1/2}+\right.\nonumber\\&+&
\left.
2T\ln\left\{1-\exp\Big[-\frac{(k^{2z}+M^2k^{2z-2})^{1/2}}{T}
\Big]\right\}
\right\}.
\eea
The first term identically reproduces the zero temperature result from
\cite{our1}.   Therefore  (throughout this paper
we adopt dimensional regularization with minimal subtraction), 
\bea
\label{u1t}
U^{(1)}&=&-\frac{d\pi^{\frac{d-1}{2}}}{{4(2\pi)^d}}
(M^2)^{\frac{d+z}{2}}\frac{\Gamma\Big(-\frac{d+z}{2}\Big)
\Gamma\Big(\frac{d+z-1}{2}\Big)}{\Gamma\Big(\frac{d}{2}\Big)}+U_T,
\nonumber\\
U_T&=&
Td \int\frac{d^dk}{(2\pi)^d}\ln(1-\exp[-\frac{(\vec{k}^{2z}+
M^2k^{2z-2})^{1/2}}{T}]).
\eea
 Notice that
$U_0$, the first term in the above expression, is finite for   $d+z$
odd while, for   $d+z$ even,
it diverges and requires a
subtraction which may be carried out by adding  a
corresponding counterterm.  Therefore, in principle, for the case
$d+z=2n$, one should introduce into the theory an additional vertex
$\lambda(\Phi\Phi^*)^n$, to achieve multiplicative renomalizability;
the presence of this new self-interaction  vertex generates new
Feynman diagrams making the evaluation of the one-loop effective
potential much more complicated. We will defer the discussion of this
situation 
to the next section and here we will restrict ourselves to the
analysis of the case with $d+z=2n+1$. By making the 
 change of variables $\frac{k^z}{T}=\bar{k}$ (with $\bar{k}$ 
 dimensionless), we obtain 
\bea
\label{u1ta}
U_T= \frac{d}{(4\pi)^{d/2}\Gamma(d/2)}T^{1+\frac{d}{z}}
\int_0^{\infty} d\bar{k}\bar{k}^{d/z-1}\ln\left[1-\exp\left(-(\bar{k}^2+
\frac{M^2}{T^{2/z}}\bar{k}^{2(1-1/z)})^{1/2}\right)\right]
\eea

Thus, for large $T$, the leading contributions are
\bea
\label{u1tb}
U_T=\frac{d}{(4\pi)^{d/2}\Gamma(d/2)}T^{1+\frac{d}{z}}\Big[A+B
\frac{M^2}{T^{2/z}}
\Big]+\ldots, 
\eea
where
\bea
\label{a}
A=\int_0^{\infty}d\bar{k} \bar{k}^{d/z-1}\ln(1-e^{-\bar{k}})={\rm
  Li}_{\frac{d}z+1}(1)
\Gamma(d/z),
\eea
and
\bea
B=\frac{1}{2}\int_0^{\infty}
\frac{d\bar{k}\bar{k}^{\frac{d-2}{z}}}{e^{\bar{k}}-1}=
\frac{1}{2}{\rm Li}_{\frac{d-2}{z}+1}(1)\Gamma(\frac{d-2}{z}+1),
\eea
where ${\rm Li}_\nu (x)$ denotes the  polylogarithm function of  order $\nu$.
We see that for $z=1$ this expression reproduces the temperature
dependence  found in \cite{DJ}, but,  for generic $z>1$, the
effective potential grows more slowly  with the temperature. 

\section{Scalar Quantum Electrodynamics with $d+z$ even}
As it was pointed out before, for the consistency of the model when
$d+z=2n$, it is necessary the inclusion of a self-interaction term for the
scalar field, so that the Lagrangian 
then becomes
\bea 
\label{lasqed1}
L&=&\frac{1}{2}F_{0i}F_{0i}+(-1)^z\frac{1}{4}F_{ij}\Delta^{z-1} F_{ij}+D_0\phi
(D_0\phi)^*
\nonumber\\
&&- D_{i_1}D_{i_2}\ldots D_{i_z}\phi(D_{i_1}D_{i_2}\ldots D_{i_z}\phi)^*-\lambda(\phi\phi^*)^n.
\eea 
The gauge fixing term and propagators for the gauge field  are the
same as in the previous section. The propagator for the scalar field will be fixed shortly.

First, we can find the contribution to the effective potential
coming from  the gauge propagators only. The corresponding Feynman
diagrams are depicted in Fig. 1.

\begin{figure}[ht]
\centerline{\includegraphics{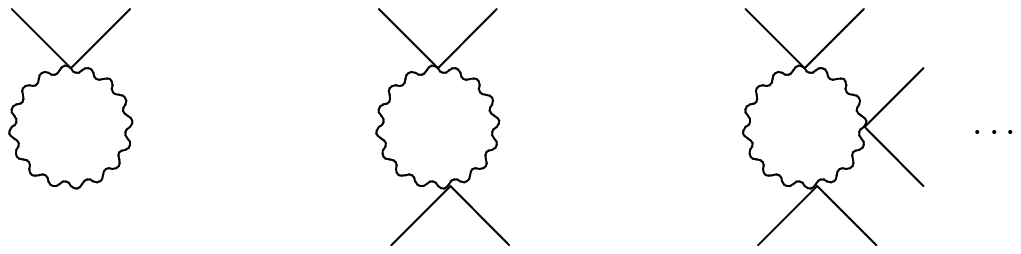}} 
\caption{Contributions involving  gauge propagators only.}
\end{figure}

There are two types of such contributions -- the first of them, $U_{a}$, is given by the
sum of loops of $<A_0A_0>$ propagators, and -- the second one, $U_{b}$,  of
loops of $<A_iA_j>$ propagators. They are completely analogous, up to
an overall factor ($U_b$ carries the factor $d$), and they
together contribute  as (cf. \cite{our,our1})
\bea
\label{uab}
U_a+U_b&=&\frac{1}{2}(d+1)\int\frac{d^{d}kdk_0}{(2\pi)^{d+1}}\ln(1+
\frac{2e^2\Phi
\Phi^*k^{2z-2}}{k^2_0+k^{2z}}).
\eea
Repeating the calculations of the Section II, one can show
that the results for $U_a+U_b$  at zero and finite temperature,
reproduce the expressions
(\ref{u1t},\ref{u1ta},\ref{u1tb}) with the only
difference that the overall factor $d$ is replaced
 by $d+1$.

Now, let us obtain the background-dependent effective propagators of
the scalar fields. After the background-quantum splitting $\phi\to\Phi+\phi$,
$\phi^*\to \Phi^*+\phi^*$, the part of the  Lagrangian
quadratic  in the quantum field $\phi$ turns out to be nontrivial,
being of the form
\bea
L_{2\phi}&=&-\phi[\pa^2_0+(-\Delta)^z]\phi^*-\nonumber\\
&-&\lambda\left\{\frac{n(n-1)}{2}(\Phi\Phi^*)^{n-2}
[(\Phi^*)^2\phi^2+\Phi^2(\phi^*)^2]+n^2\Phi^{n-1}(\Phi^*)^{n-1}\phi\phi^*
\right\},
\eea
which generates the propagators for $\phi$: 
\bea
\label{prophi}
& &\left(
\begin{array}{cc}
<\phi\phi> & <\phi\phi^*>\\
<\phi^*\phi> & <\phi^*\phi^*>
\end{array}
\right)=
i\left(
\begin{array}{cc}
{\cal M} & \pa^2_0+(-\Delta)^z+\mu\\
\pa^2_0+(-\Delta)^z+\mu & \bar{\cal M}
\end{array}
\right)^{-1}=
\\
&=&\frac{i}{(\pa^2_0+(-\Delta)^z+\mu)^2-{\cal M}\bar{\cal M}}\left(
\begin{array}{cc}
\bar{\cal M} & -(\pa^2_0+(-\Delta)^z+\mu)\\
-(\pa^2_0+(-\Delta)^z+\mu) & {\cal M}
\end{array}
\right),\nonumber
\eea
where ${\cal M}=\lambda n(n-1)(\Phi\Phi^*)^{n-2}(\Phi^*)^2$,
$\bar{\cal M}=\lambda n(n-1)(\Phi\Phi^*)^{n-2}\Phi^2$, and $\mu=
\lambda n^2(\Phi\Phi^*)^{n-1}$.
These propagators will be represented by bold
straight lines.

Besides, we also must use the background-dependent propagators of the
gauge field, $<A_0A_0>$ and $<A_iA_j>$, which are introduced as a
result of the following summation over the quartic vertices
represented in 
Fig. 2:

\begin{figure}[ht]
\centerline{\includegraphics{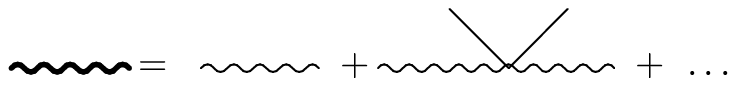}} 
\caption{Background-dependent gauge propagator.}
\end{figure}

Actually, we will use not these propagators themselves but the objects
derived from them:
\bea
G_1&=&<\partial_0A_0\partial_0A_0>;
\nonumber\\
G_2&=&<\partial_i\Delta^{z-1}A_i\partial_j\Delta^{z-1}A_j>,
\eea
whose Fourier transforms in the Euclidean space are 
\bea
\label{efpr}
G_1(k)&=&\frac{k^2_0\vec{k}^{2z-2}}{k^2_0+
\vec{k}^{2z}+2e^2\vec{k}^{2z-2}
\Phi\Phi^*};
\nonumber\\
G_2(k)&=&\frac{
\vec{k}^{4z-2}}{k^2_0+\vec{k}^{2z}+2e^2\vec{k}^{2z-2}\Phi\Phi^*}.
\eea

The presence of the new propagators will lead to a
contribution to the effective action emerged from the "crossed" sector
(those graphs involving both gauge and matter propagators shown at
Fig. 3) given by

\begin{figure}[ht]
\centerline{\includegraphics{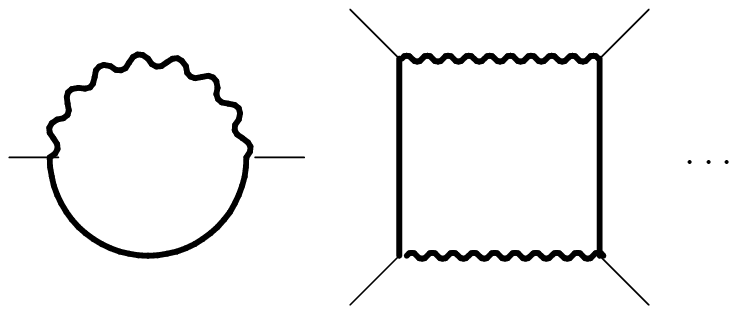}} 
\caption{Contributions involving  gauge and matter propagators.}
\end{figure}

\bea
\label{sumgc1}
U_c&=&-\frac{1}{2}\sum\limits_{n=1}^{\infty}\frac{1}{n}\int
\frac{d^dkdk_0}{(2\pi)^{d+1}}\Big[-e^2(2\Phi\Phi^*<\phi\phi^*>+\Phi\Phi
<\phi^*\phi^*>+\Phi^*\Phi^*<\phi\phi>)\times\nonumber\\&\times&
(G_1+G_2)\Big]^n,
\eea

where
\bea
&& 2\Phi\Phi^*<\phi\phi^*>+\Phi\Phi<\phi^*\phi^*>+\Phi^*\Phi^*<\phi\phi>=
\nonumber\\ &=&
-\frac{2\Phi\Phi^*(k^2_0+k^{2z}+\mu)+{\cal M}\Phi\Phi+\bar{\cal
    M}\Phi^*\Phi^*}{(k^2_0+
k^{2z}+\mu)^2-{\cal M}\bar{\cal M}}.
\eea

So, we can write
\bea
\label{gc1}
U_c&=&-\frac{1}{2}\sum\limits_{n=1}^{\infty}\frac{1}{n}\int
\frac{d^dkdk_0}{(2\pi)^{d+1}}
\Big(e^2
\frac{2\Phi\Phi^*(k^2_0+k^{2z}+\mu)+{\cal M}\Phi\Phi+\bar{\cal M}\Phi^*\Phi^*}
{(k^2_0+k^{2z}+\mu)^2-{\cal M}\bar{\cal M}}\times\nonumber\\&\times&
\cdot
\frac{k^2_0\vec{k}^{2z-2}+k^{4z-2}}{k^2_0+\vec{k}^{2z}+2e^2\vec{k}^{2z-2}
\Phi\Phi^*}\Big)^n=\nonumber\\
&=&\frac{1}{2}\int
\frac{d^dkdk_0}{(2\pi)^{d+1}}
\ln \Big(1-e^2 \frac{2\Phi\Phi^*(k^2_0+k^{2z}+\mu)+{\cal M}\Phi\Phi+
\bar{\cal M}\Phi^*\Phi^*}{(k^2_0+k^{2z}+\mu-
\sqrt{{\cal M}\bar{\cal M}})(k^2_0+k^{2z}+\mu+\sqrt{{\cal M}\bar{\cal
    M}})}\times
\nonumber\\&\times&
\frac{k^{2z-2}(k^2_0+k^{2z})}{k^2_0+\vec{k}^{2z}+
M^2k^{2z-2}}\Big).
\eea

The integral over momenta, as well as the discretization of the zero 
component of the
momentum in order to implement finite temperature, is
straightforward but
the result is highly cumbersome. Nevertheless it can be performed in the following way.

To simplify this expression, we introduce $\mu_{\pm}=\mu\pm\sqrt{{\cal M}\bar{\cal M}}$, so that $\mu_+=\lambda n(2n-1)(\Phi\Phi^*)^{n-1}$, and $\mu_-=\lambda n(\Phi\Phi^*)^{n-1}$ and define $q^2=k^2_0+k^{2z}$.Using that  ${\cal M}\Phi\Phi+\bar{\cal M}\Phi^*\Phi^*=2\lambda n(n-1)(\Phi\Phi^*)^n$  we get
\bea
U_c&=&\frac{1}{2}
\int
\frac{d^dkdk_0}{(2\pi)^{d+1}}
\ln \Big(1-e^2 \frac{2\Phi\Phi^*(q^2+\mu)+2\lambda n(n-1)(\Phi\Phi^*)^n}{(q^2+\mu_-)(q^2+\mu_+)}
\frac{k^{2z-2}q^2}{q^2+
M^2k^{2z-2}}\Big).
\eea
After some algebraic transformations, we can rewrite this expression as
\bea
U_c&=&\frac{1}{2}
\int
\frac{d^dkdk_0}{(2\pi)^{d+1}}
\Big[\ln \Big((q^2+\mu_+)(q^2+\mu_-)(q^2+k^{2z-2}M^2)-q^2k^{2z-2}M^2(q^2+\mu_+)\Big)-\nonumber\\&-&
\ln(q^2+\mu_+)-\ln(q^2+\mu_-)-\ln(q^2+k^{2z-2}M^2)\Big].
\eea
By cancelling  the term with $\ln(q^2+\mu_+)$, we arrive at
\bea
U_c&=&\frac{1}{2}
\int
\frac{d^dkdk_0}{(2\pi)^{d+1}}
\Big[\ln \Big((q^2+\mu_-)(q^2+k^{2z-2}M^2)-q^2k^{2z-2}M^2\Big)-\nonumber\\&-&
\ln(q^2+\mu_-)-\ln(q^2+k^{2z-2}M^2)\Big].
\eea
It follows from (\ref{uab}) that $U_a=\frac{1}{2}\int\frac{d^dkdk_0}{(2\pi)^{d+1}}\ln(q^2+M^2k^{2z-2})$. Thus, we can cancel some  additional terms and get
\bea
\label{uabc}
U_a+U_c&=&\frac{1}{2}
\int
\frac{d^dkdk_0}{(2\pi)^{d+1}}
\Big[\ln \Big((q^2+\mu_-)(q^2+k^{2z-2}M^2)-q^2k^{2z-2}M^2\Big)- 
\ln(q^2+\mu_-)\Big];\nonumber\\
U_b&=&\frac{d}{2}\int\frac{d^dkdk_0}{(2\pi)^{d+1}}\ln(q^2+M^2k^{2z-2}).
\eea

To close the calculations, let us obtain the contribution to the effective potential
generated by the self-coupling of the scalar fields.
It follows from (\ref{prophi}) that in this case we have
a new contribution to the effective action
\bea
U^{(1)}_d=-\frac{i}{2}\ln\det\left(
\begin{array}{cc}
{\cal M} & \pa^2_0+(-\Delta)^z+\mu\\
\pa^2_0+(-\Delta)^z+\mu & \bar{\cal M}
\end{array}
\right).
\eea
To calculate this determinant it is convenient to perform the
Fourier transform, which after a Wick rotation to the Euclidean space yields
\bea
U^{(1)}_d=\frac{1}{2}\int\frac{d^dkdk_0}{(2\pi)^{d+1}}\ln\det\left(
\begin{array}{cc}
{\cal M} & k^2_0+k^{2z}+\mu\\
k^2_0+k^{2z}+\mu & \bar{\cal M}
\end{array}
\right).
\eea

Up to an irrelevant additive constant, the evaluation of this expression gives
\bea
&&U^{(1)}_d=\frac{1}{2}\int\frac{d^dkdk_0}{(2\pi)^{d+1}}\ln\Big[
(k^2_0+k^{2z}+\mu)^2-{\cal M}\bar{\cal M}\Big]=
\nonumber\\
&=&
\frac{1}{2}\int\frac{d^dkdk_0}{(2\pi)^{d+1}}\Big(\ln[q^{2}+\mu_-
]+\ln[q^{2}+\mu_+]\Big),
\eea

Using (\ref{uabc}), we can write the complete one-loop effective potential:
\bea
\label{uabcd}
U_a+U_b+U_c+U_d&=&\frac{1}{2}
\int
\frac{d^dkdk_0}{(2\pi)^{d+1}}
\Big[\ln \Big((q^2+\mu_-)(q^2+k^{2z-2}M^2)-q^2k^{2z-2}M^2\Big)
+\nonumber\\
&+&d\ln(q^2+M^2k^{2z-2})+\ln(q^2+\mu_+)\Big].
\eea
The term proportional to $d$ (that is, $U_b$) is given by the expressions (\ref{u1t}--\ref{u1tb}). The last term, that is, those originated from $U_d$ involving $\ln(q^2+\mu_+)$, yields the zero temperature result
\bea
\label{zerotemp}
U^{(1)}_{d+}&=&-\frac{1}{2\sqrt{\pi}}\frac{1}{(2\pi)^d}\frac{1}{z}
\frac{\pi^{d/2}}{\Gamma(d/2)}\Gamma(\frac{d}{2z})\Gamma(-\frac{1}{2}-
\frac{d}{2z})
\mu_+^{1/2+d/(2z)},
\eea
plus the finite temperature contribution
\bea
U^{(1)}_{d+}(T)&=&2T\int\frac{d^dk}{(2\pi)^d}\left\{
\ln\left\{1-\exp\left[-\Big(\frac{\sqrt{k^{2z}+\mu_+}}{T}
\Big)\right]\right\}
\right\}.
\eea
At high temperatures we obtain
\bea
\label{htemp}
U^{(1)}_{d+}(T)&=&
\frac{2}{(4\pi)^{d/2}\Gamma(d/2)}T^{1+\frac{d}{z}}(A+
\mu_+\frac{B_0}{T^2})+\ldots,
\eea
where $A$ is given by (\ref{a}), and
\bea
B_0&=&\int_0^{\infty}d\bar{k}
\frac{\bar{k}^{d/z-2}}{e^{\bar{k}}-1}={\rm Li}_{d/z-1}(1)\Gamma(d/z-1).
\eea
It remains to analyse  the first term from (\ref{uabcd}) which looks like
\bea
I=\frac{1}{2}
\int
\frac{d^dkdk_0}{(2\pi)^{d+1}}
\ln \Big((q^2+\mu_-)(q^2+k^{2z-2}M^2)-q^2k^{2z-2}M^2\Big).
\eea
Unfortunately, this integral cannot be done in a closed form.  We present the results only for two particular cases.

(i) When the contribution of the gauge coupling dominates,  we can choose $\lambda\simeq 0$. Then one has $\mu_-= 0$ and the integral $I$ is just an irrelevant constant, independent of the classical fields. The complete contribution to the effective potential in this case comes from the terms (\ref{u1t}--\ref{u1tb}), while the term proportional to $\lambda$ is essential only on the tree level.

(ii) When the contribution of the scalar self-coupling dominates, we can choose $g\simeq 0$ in this term. In this case, one has $M=0$, so, $I\simeq \frac{1}{2}\int\frac{d^dkdk_0}{(2\pi)^{d+1}}\ln (q^2+\mu_-)$, which, similarly to expressions (\ref{zerotemp}) and (\ref{htemp}), yields
\bea
\label{zerotemp1}
U^{(1)}_{d-}&=&-\frac{1}{2\sqrt{\pi}}\frac{1}{(2\pi)^d}\frac{1}{z}
\frac{\pi^{d/2}}{\Gamma(d/2)}\Gamma(\frac{d}{2z})\Gamma(-\frac{1}{2}-
\frac{d}{2z})
\mu_-^{1/2+d/(2z)},
\eea
at zero temperature and

\bea
\label{htemp1}
U^{(1)}_{d+}(T)&=&
\frac{2}{(4\pi)^{d/2}\Gamma(d/2)}T^{1+\frac{d}{z}}(A+
\mu_+\frac{B_0}{T^2})+\ldots.
\eea
at high temperature.
Here the effective potential is reduced to the sum of the expressions (\ref{zerotemp}) and (\ref{zerotemp1}) at zero temperature, and of  (\ref{htemp}) and (\ref{htemp1}) at the high temperature.

We close this section with a discussion of the renormalizability of
the model. First, one reminds that we introduced  a self-coupling of the
scalar field since, at $\lambda=0$, the one-loop effective
potential diverges if $d+z=2n$ with $n$ integer (see (\ref{u1t})) so, 
the counterterms 
$(\Phi\Phi^*)^n$ is needed. At the same time, the new vertex 
$(\Phi\Phi^*)^n$ generates new contributions as in (\ref{zerotemp}), 
which diverge if
$1+\frac{d}{z}=2\tilde{n}$, with $\tilde{n}$ an integer. These
contributions are proportional to
$(\Phi\Phi^*)^{\tilde{n}(n-1)}$, therefore, to achieve multiplicative
renormalizability, one must, in principle, introduce a new vertex
$(\Phi\Phi^*)^{\tilde{n}(n-1)}$, which, again modifies the classical
action. The only exceptional situation, when this modification is not
necessary, is the case $\tilde{n}=\frac{n}{n-1}$. For $n$ and
$\tilde{n}$ integer, the only solution is $n=\tilde{n}=2$. Therefore,
we conclude that only the vertex $(\Phi\Phi^*)$
corresponds to the renormalizable interaction, with $d=3$ and $z=1$,
that is, just the usual scalar QED. We note, however, that in the
cases, where either $d+z$ is odd (that is, the case considered in the
previous section), or $\frac{d}{z}$ is not an odd
number (i.e. either even, or fractionary one), this problem simply
will not arise, since, for $d+z$ odd, there is no divergent
contributions to the one-loop effective potential.
 There are, of course, additional restrictions on $d$
and $z$ arising from the fact that, in the renormalizable theories,
dimensions of couplings must be non-negative, i.e. $z-d+2\geq 0$ (for the
coupling $e$) and $d+z-n(d-z)\geq 0$ (for the coupling
$\lambda$). However, these restrictions play a role only at higher
loop orders.

\section{Yukawa theory}

Let us now formulate the arbitrary $z$ version of the Yukawa theory whose
Lagrangian density
is
\bea
\label{yukawa}
L=\bar{\psi}(i\gamma^0\pa_0+(i\gamma^i\partial_i)^z+h\Phi)\psi.
\eea
To study the one-loop effective potential, it is enough to treat 
the scalar field
as  purely external, and to consider the spinor field to be massless
since a nontrivial mass implies only in a redefinition of the
$\Phi$ field. In this case, the loop expansion ends at the
one-loop contribution. 
However, if we assume that  $\Phi$ is also dynamical (which, in
particular, is necessary to proceed renormalization if 
the contribution to the effective potential diverges),
with the same critical exponent $z$ as the $\psi$, its free
Lagrangian is the same as in the theory (\ref{lasqed}). Notice that the mass
dimension of $h$ is $(3z-d)/2$, and the theory is
(super)renormalizable for $z\geq d/3$
-- in
particular, it is renormalizable in the   usual case ($z=1$) Yukawa model in  
$(3+1)$-dimensional space. 

The one-loop effective potential corresponding to the Lagrangian
(\ref{yukawa}),  looks like
\bea
U^{(1)}=i {\rm Tr}\ln(i\gamma^0\pa_0+(i\gamma^i\partial_i)^z+h\Phi).
\eea
We have two possibilities. In the first one, $z$ is even, so, 
$(i\gamma^i\partial_i)^z=(-\Delta)^{z/2}$, and
we find that the effective potential in the Euclidean space is
\bea
U^{(1)}=-\frac{\delta}{2}\int\frac{d^dkdk_0}{(2\pi)^{d+1}}
\ln\Big[\frac{k^2_0+(k^2)^{z/2}+h\Phi)^2}{k^2_0}
\Big],
\eea
where $\delta$ is the dimension of the Dirac matrices.
Taking into account the discretization of the
zero 
component of the momentum, we get
\bea
U^{(1)}=-T\frac{\delta}{2}\sum\limits_{n=-\infty}^{\infty}
\int\frac{d^dk}{(2\pi)^d}
\ln\Big[4\pi^2T^2(n+\frac{1}{2})^2+((k^2)^{z/2}+h\Phi)^2
\Big].
\eea
Using the expression for the sum 
\bea
\label{sumferm}
\sum\limits_{n=-\infty}^{\infty}
\ln(\pi^2(2n+1)^2T^2+E^2)=\frac{E}{T}+2\ln(1+e^{-E/T}).
\eea
(cf. \cite{DJ}; note that the presence of
$\ln(1+\cdots)$ instead of $\ln(1-\cdots)$ comes from a difference between
bosonic and fermionic  cases), we arrive at
\bea
U^{(1)}=-\frac{\delta}{2}\int\frac{d^dk}{(2\pi)^d}\Big[(k^2)^{z/2}+
h\Phi+2T\ln(1+\exp[-\frac{(k^2)^{z/2}+h\Phi}{T}])
\Big].
\eea
Integration of the first term gives zero result, and hence, no 
renormalization is
needed. So, we 
get
\bea
U^{(1)}=-T\delta\int\frac{d^dk}{(2\pi)^d}\ln(1+\exp[-\frac{(k^2)^{z/2}
+h\Phi}{T}]).
\eea
This expression is non-trivial only for  non-zero temperature. Proceeding
just as 
in the previous sections, we find at  large $T$:
\bea
U^{(1)}=-\frac{T^{1+d/z}\delta}{2^d\pi^{d/2}\Gamma(d/2)z}(A_1-
\frac{B_1h\Phi}{T}),
\eea
where
\bea
\label{a1}
A_1&=&\int_0^{\infty}d\bar{k}
\bar{k}^{d/z-1}\ln(1+e^{-\bar{k}})=(-1+2^{d/z})
\Gamma(\frac{d}{z})\zeta(\frac{d+z}{z})2^{-d/z};\nonumber\\
B_1&=&\int_0^{\infty}\frac{d\bar{k} \bar{k}^{d/z-1}}{e^{\bar{k}}+1}= 
(2^{d/z}-2)\Gamma(\frac{d}{z})\zeta(\frac{d}{z})2^{-d/z},
\eea
where $\zeta(x)$ is the Riemann zeta function.

The second possibility when $z$ is odd, $z=2l+1$, so that, 
$(i\gamma^i\partial_i)^z=(-\Delta)^l i\gamma^i\partial_i$. In this case we have
\bea
U^{(1)}=
-i{\rm Tr}\ln(i\gamma^0\pa_0+i(-\Delta)^l  \gamma^i\partial_i+h\Phi)=
-\frac{1}{2}\delta\int\frac{dk_0d^dk}{(2\pi)^{d+1}}
\ln(k^2_0+(k^2)^z+M^2),
\eea
where $M=h\Phi$.
We replace the zero component of the momentum by the discrete one,
$k_0=(2n+1)\pi T$, and have
\bea
U^{(1)}=-\frac{1}{2}\delta T\sum\limits_{n=-\infty}^{\infty}
\int\frac{d^dk}{(2\pi)^d}\ln(4\pi^2T^2(n+\frac{1}{2})^2+(k^2)^z+M^2)
\eea
After evaluating  the sum we get
\bea
U^{(1)}=-\frac{1}{2}\delta
\int\frac{d^dk}{(2\pi)^d}\Big[((k^2)^z+M^2)^{1/2}+
T\ln(1+\exp[-\frac{(k^{2z}+M^2)^{1/2}}{T}])
\Big].
\eea
Performing the integration, we find
\bea
\label{u1spi}
U^{(1)}&=&U_0+U_T=\nonumber\\
&=&-\delta\frac{\pi^{d/2-1/2}}{(2\pi)^d(d+z)}\frac{1}{\Gamma(d/2)}
\Gamma\Big(\frac{d}{2z}\Big)
\Gamma\Big(\frac{1}{2}\big[1-\frac{d}{z}\big]\Big)(h\Phi)^{\frac{d}{z}+1}-
\nonumber\\&-&
\frac{1}{2}T\delta
\int\frac{d^dk}{(2\pi)^d}\ln(1+\exp[-\frac{(k^{2z}+
M^2)^{1/2}}{T}]).
\eea
As before, we have a sum of the zero temperature result with an
 additive term which is non-trivial only at the non-zero
temperature. { As for the zero temperature term, it is proportional
  to $\Gamma\Big(\frac{1}{2}\big[1-\frac{d}{z}\big]\Big)$, thus, it
  diverges if $1-\frac{d}{z}=-2n$, with $n$ integer. In this case the
  divergence will be proportional to $\Phi^{2n+2}$. Thus, similarly to the
  previous section, the
  $\Phi^{2n+2}$ term must be present in the Lagrangian from the very
  beginning. Its presence will give an additional contribution to the
  effective action. Such a contribution, at the one-loop level, is a
  sum of all one-loop scalar graphs. Therefore it enters the one-loop
  effective action only as an additive term. Indeed, if the action of the
  scalar field looks  like
\bea
S=\int dt d^dx
(\frac{1}{2}\dot{\phi}^2-\frac{1}{2}(-1)^z\phi\Delta^z\phi-
V(\phi)),
\eea
the corresponding one-loop effective potential, in the case
$1-\frac{d}{z}=-2n$,
is \cite{our}
\bea
U^{(1)}=-\frac{1}{2\sqrt{\pi}}\frac{1}{(2\pi)^d}\frac{1}{z}
\frac{\pi^{d/2}}{\Gamma(d/2)}\Gamma(n+\frac{1}{2})\Gamma(-1-n)
(V^{\prime\prime}(\Phi))^{n+1}.
\eea
with $\Phi$ being a background field, and
$V(\Phi)=\frac{f}{(2n+2)(2n+1)}\Phi^{2n+2}$, so that,
$V^{\prime\prime}(\Phi)=f\Phi^{2n}$. The corresponding quantum
correction, for $1-\frac{d}{z}=-2n$, with $n$ integer, is divergent
being proportional to $\Gamma(-n-1)\Phi^{2n(n+1)}$, which, of course,
needs a  counterterm, and, hence, the presence of this
vertex in the action from the very beginning, which, consequently,
modifies $V(\Phi)$ once more. The only special case is $n=1$, where
this modification does not happen, it corresponds to 
$d=3z$. This
divergent term reproduces the structure of the potential, i.e. if the
coupling looks like $V(\Phi)=\frac{f}{12}\Phi^4$, the divergences
arising both from spinor and scalar sectors will be proportional to
$\Phi^4$, so, no other coupling
 is needed in this  case. Actually, we
have shown that it is the only renormalizable case.}

The temperature dependent term from (\ref{u1spi}), after the
corresponding  change of variables, is
\bea
U_T=-\frac{T^{1+d/z}\delta}{2^{d+1}\pi^{d/2}\Gamma(d/2)z}\int_0^{\infty}
d\bar{k } \bar{k}^{d/z-1}\ln(1+e^{-\sqrt{\bar{k}^2+M^2/T^2}}).
\eea
Again, we can obtain leading and subleading terms: 
\bea
U_T=-\frac{T^{1+d/z}\delta}{2^{d+1}\pi^{d/2}\Gamma(d/2)z}(A_1-B_2
\frac{M^2}{T^2}),
\eea
where
\bea
B_2=\frac{1}{2}\int_0^{\infty}\frac{d\bar{k}
  \bar{k}^{d/z-2}}{e^{\bar{k}}+1}=
(2^{d/z}-4)\Gamma(\frac{d}{z}-1)\zeta(\frac{d}{z}-1)2^{-(d+z)/z},
\eea
and $A_1$ is just the same one defined earlier in (\ref{a1}). We note
that this integral is well  defined if $d>z$.
Thus, we obtained the high-temperature asymptotic expressions for the
effective potential,  both in bosonic and fermionic case.

\section{conclusions}

In this work we studied the one-loop effective potential at finite temperature
for the HL QED and Yukawa models. We made also important remarks on the zero
temperature situation which extend an earlier study by some of us.
Indeed, for $d+z$ even in the case of QED and also for the Yukawa model with $z$
odd there occurs divergences whose elimination require the addition of
self-interactions
of the scalar fields. These new terms produce new divergences in a way
that  invalidates
the usual renormalization procedure unless for the usual case, $z=1$ and $d=3$.
In the cases with $d+z$ odd for the HL QED and $z$ even for the HL Yukawa models there
are no one-loop divergences. This, of course, does not preclude the existence of divergences
in higher orders which must be removed by an adequate renormalization scheme.

The models have the same  high temperature  limit proportional to  $T^{1+d/z}$ 
as they should but different next to leading behaviors unless $d=z$. Thus for $z>1$ the efffective potential
decays  with the temperature more slowly than in the usual case \cite{DJ}.

{\bf Acknowledgements.} This work was partially supported by Conselho
Nacional de Desenvolvimento Cient\'{\i}fico e Tecnol\'{o}gico (CNPq)
and Funda\c{c}\~{a}o de Amparo \'{a} Pesquisa de Estado de S\~{a}o
Paulo  (FAPESP).
A. Yu. P. has been supported by the CNPq project 303438-2012/6.

\end{document}